\documentclass[runningheads]{llncs}
\usepackage{graphicx}
\usepackage{array}
\usepackage{mathtools}
\usepackage[draft]{fixme}
\usepackage{xcolor}
\usepackage{tikz}
\usetikzlibrary{hobby,positioning, shapes,patterns,shadows.blur, arrows,decorations,arrows,automata,shadows,patterns,chains,graphs,calc,intersections,matrix,fit,shapes,chains,decorations.pathreplacing}
\usepackage{multicol}
\usepackage{listings}
\usepackage{comment}
\usepackage{dsfont}
\usepackage{url}
\usepackage{wrapfig}
\usepackage{comment}
\usepackage{paralist}
\usepackage{fontawesome5}

\fxsetup{theme=colorsig,layout=inline}

\definecolor{keywords}{HTML}{f9005a}
\definecolor{identifiers}{HTML}{333333}
\definecolor{comments}{HTML}{75715E}
\definecolor{typecol}{HTML}{679c00}

\lstdefinestyle{custompython}{
  belowcaptionskip=0pt,
  breaklines=true,
  frame=none,
  xleftmargin=1em,
  xrightmargin=-0.5cm,
  numbersep=2pt,
 language=python,
  showstringspaces=false,
  basicstyle=\scriptsize\ttfamily,
  keywordstyle=\bfseries\color{keywords},
  commentstyle=\itshape\color{comments},
  identifierstyle=\color{identifiers},
  stringstyle=\color{orange},
  numberstyle=\tiny\color{gray},
  numbers=left, 
  breaklines=true,
  postbreak=\mbox{{$\hookrightarrow$}\space},
}

\newcommand{\pythonstyle}[2]{\lstset{
  belowcaptionskip=0pt,
  breaklines=true,
  frame=none,
  xleftmargin=1em,
  xrightmargin=-0.5cm,
  numbersep=2pt,
 language=python,
  showstringspaces=false,
  basicstyle=\footnotesize\ttfamily,
  keywordstyle=\bfseries\sffamily\color{black!90},
  commentstyle=\itshape\color{comments},
  identifierstyle=\color{identifiers},
  stringstyle=\color{orange},
  numberstyle=\tiny\color{gray},
  numbers=left, 
  breaklines=true,
  postbreak=,
  morekeywords = {Select,Extract,From,Where,Apply},
  mathescape = true,
}}

\usetikzlibrary{decorations.pathreplacing,calligraphy}

\lstnewenvironment{pythonenv}[2]{\pythonstyle{#1}{#2}}{}

\newcommand\qline[1]{{\pythonstyle{0}{}\lstinline!#1!}}

\usepackage{etoolbox}
\usepackage{comment}
\usepackage{braket}
\usepackage{hyperref}
\usepackage{cite}
\usepackage[figurename=Figure]{caption}

\definecolor{mygreen}{HTML}{666666}
\definecolor{mygray}{HTML}{666666}
\definecolor{mymauve}{HTML}{666666}
\definecolor{terminalbgcolor}{HTML}{333333}
\definecolor{terminalrulecolor}{HTML}{000000}
\newcommand{\lstconsolestyle}{
\lstset{
  backgroundcolor=\color{black!2},
  basicstyle=\color{black}\fontfamily{fvm}\tiny\selectfont,
  breakatwhitespace=false,  
  breaklines=true,
  captionpos=b,
  commentstyle=\color{mygreen},
  deletekeywords={...},
  escapeinside={\%*}{*)},
  extendedchars=true,
  frame=single,
  keepspaces=true,
  keywordstyle=\color{mygray},
  morekeywords={*,...},
  numbers=none,
  numbersep=5pt,
  framerule=1pt,
  numberstyle=\color{mygray}\tiny\selectfont,
  rulecolor=\color{terminalrulecolor},
  showspaces=false,
  showstringspaces=false,
  showtabs=false,
  stepnumber=2,
  stringstyle=\color{mymauve},
  tabsize=2
}
}

\newcommand{\lstconsolestyleX}{
\lstset{
  backgroundcolor=\color{black!2},
  basicstyle=\color{black}\fontfamily{fvm}\footnotesize\selectfont,
  breakatwhitespace=false,  
  breaklines=true,
  captionpos=b,
  commentstyle=\color{mygreen},
  deletekeywords={...},
  escapeinside={\%*}{*)},
  extendedchars=true,
  frame=single,
  keepspaces=true,
  keywordstyle=\color{black},
  morekeywords={*,...},
  numbers=none,
  numbersep=0pt,
  framerule=0pt,
  numberstyle=\color{black}\tiny\selectfont,
  rulecolor=\color{black},
  showspaces=false,
  showstringspaces=false,
  showtabs=false,
  stepnumber=0,
  stringstyle=\color{black},
  tabsize=1
}
}

\newcommand\tline[1]{{\lstconsolestyleX\lstinline!#1!}}
\newcommand{\toolname}{\textsc{SMTQuery}}
\newcommand{\qlang}{\textsc{qlang}}
\newcommand{\zthree}{\textsc{Z3}}
\newcommand{\zstr}{\textsc{Z3Str3}}
\newcommand{\cvc}{\textsc{cvc5}}
\newcommand{\vsep}{\hspace{3mm} | \hspace{3mm}}
\newcommand{\woorpje}{\textsc{Woorpje}}
\newcommand{\zaligVinder}{\textsc{ZaligVinder}}
\newcommand{\toolWeb}{\url{http://smtquery.github.io}}

\newcommand{\citep}[1]{\cite{#1}}

\begin{document}
\title{A Generic Information Extraction System for String Constraints}
\author{Joel D. Day\inst{1}, Adrian Kr\"oger\inst{2}, Mitja~Kulczynski\inst{3}, Florin~Manea\inst{2}, Dirk~Nowotka\inst{3} and Danny~B\o gsted~Poulsen\inst{4}}
\authorrunning{J. Day, A. Kr\"oger, M. Kulczynski, F. Manea, D. Nowotka and D. Poulsen}
\institute{
	Department of Computer Science, Loughborough University,
	Loughborough, UK
	\and
	Department of Computer Science, University of Göttingen, Göttingen,
	Germany
	\and
	Department of Computer Science, Kiel University, Kiel,
	Germany
	\and
	Department of Computer Science, Aalborg University, Aalborg,
	Denmark
    }
\maketitle 
\begin{abstract}
String constraint solving, and the underlying theory of word equations, are highly interesting research topics both for practitioners and theoreticians working in the wide area of satisfiability modulo theories. 
As string constraint solving algorithms, a.k.a. string solvers, gained a more prominent role in the formal analysis of string-heavy programs, especially in connection to symbolic code execution and security protocol verification, we can witness an ever-growing number of benchmarks collecting string solving instances from real-world applications as well as an ever-growing need for more efficient and reliable solvers, especially for the aforementioned real-world instances. Thus, it seems that the string solving area (and the developers, theoreticians, and end-users active in it) could greatly benefit from a better understanding and processing of the existing string solving benchmarks. In this context, we propose \toolname{}: an SMT-LIB benchmark analysis tool for string constraints. \toolname{} is implemented in \textsc{Python 3}, and offers a collection of analysis and information extraction tools for a comprehensive data base of string benchmarks (presented in SMT-LIB format), based on an SQL-centred language called \qlang{}. 
\end{abstract}

\section{Introduction}
The theory of string solving is a research area in which one is interested in the mathematical and algorithmic properties of systems of constraints involving (but not restricted to) string variables and string constants. As such, string solving is part of the general constraint satisfiability topic, where one is interested in the satisfiability of formulae modulo logical theories over strings. Recent motivations for theoretical and practical investigations in this area come from the verification of security protocols (e.g., detecting security flaws exploited in \emph{injection attacks} or \emph{cross site scripting attacks}) or the symbolic execution of string-heavy languages. Excellent overviews of the main definitions and fundamental results as well as of the many recent developments related to the theory and practice of string solving are \cite{Amadini2020,Hague19}.\looseness=-1 

Relevant to our work, on the practical side, a series of dedicated string constraint solvers were developed (see, e.g., \textsc{Norn}~\citep{abdulla2015}, \textsc{Stranger}~\citep{yu2010}, \textsc{ABC}~\citep{aydin2015}, \woorpje~\cite{woorpjesat,woorpjelevi}, \textsc{OSTRICH}~\cite{ChenHLRW19}, \textsc{CertiStr}~\cite{Kan21}), but also well-established general-purpose SMT solvers (such as \textsc{CVC5}~\citep{barbosa2022cvc5} and \textsc{Z3}~\citep{de2008z3,berzish2017,z3str4}) started offering integrated string solving components. The efforts dedicated to improving the performance of many of these solvers are still ongoing. 

Thus, having a reliable and curated collection of benchmarks containing string constraints seems to be of foremost importance for the development and evaluation of string solvers. The main benchmarks used in the evaluation of string solvers are presented in detail in \cite{zaligVinderJournal}. These benchmarks were extracted both from real-world and artificial scenarios. Some benchmarks based on real-world scenarios are related to, e.g., symbolic execution of string-heavy programs (\textsc{Kaluza}, \textsc{PyEx}, \textsc{LeetCode}), software verification (\textsc{Norn}), sanitization (\textsc{PISA}), or to detection of software vulnerabilities caused by improper string manipulation (\textsc{AppScan}, \textsc{JOACO}, \textsc{Stranger}). Some artificially produced benchmarks are based either on theoretical insights (\textsc{Sloth}, \woorpje, \textsc{Light Trau}) or on fuzzing algorithms (\textsc{BanditFuzz}, \textsc{StringFuzz}). 

\begin{lstlisting}[caption={Instance taken from the PISA set},label=lst:smtexample,captionpos=b]
(set-logic QF_SLIA)
(declare-fun v1 () String)
(declare-fun v2 () String)
(declare-fun v3 () Int)
(declare-fun ret () String)
(assert (= v2 "<") )
(assert (ite (str.contains v1 v2) (and (= v3 (str.indexof v1 v2 0)) (= ret (str.substr v1 0 v3))) (= ret v1)))
(assert (or (str.contains ret "<")  (str.contains ret ">")))
(check-sat)
\end{lstlisting}
In general, all these benchmarks contain systems of string constraints. For instance, in Listing~\ref{lst:smtexample} we depict an instance from the PISA set~\cite{zheng2013z3}. It models a conditional choice of a return string variable \texttt{ret} making assumptions on other variables having the data type String. For examples of systems of string constraints, as found in the benchmarks, see, e.g.,~\cite{barrett2015smt,smtbenchmarks}. 

Moreover, there exists now a unified string-logic standard as part of SMT-LIB, and the tool \zaligVinder~\cite{zaligVinderJournal} brings together a set of relevant benchmarks and introduces a uniform benchmarking framework. Nevertheless, there are still some notable challenges related to string solving benchmarks. Firstly, they are mostly uncategorized with respect to the type of string constraints they contain, and solvers addressing specific types of constraints have to first preprocess the existing benchmarks and extract the relevant constraints (see, for instance, the approaches and discussion in \cite{cav21,Chen2021}). Secondly, the performance of solvers on the benchmarks was sometimes hard to observe and compare: the sat or unsat answers provided by some existing tools were sometimes wrong on a relatively large set of inputs \cite{Kulczynski_2022} and the size of benchmarks means it is challenging to get a precise image on where one solver outperforms others and, perhaps even more importantly, why. \looseness=-1

\paragraph{Research Tasks.}
In this context, we can formulate a series of research tasks addressing the main issues related to curating and processing string solving benchmarks. 
\begin{compactenum}
\item Identify, store, and organize a comprehensive collection of benchmarks for string solving as a database, allowing querying, exporting, and information extraction from the benchmarks, as well as an interface for running supported string solvers on specific benchmarks, extracted w.r.t. certain requirements from the entire database, and evaluating their performance. 
\item Offer functionalities allowing the extension of the database with new benchmarks, as well as the integration of new string solvers. 
\end{compactenum}

\noindent A tool answering these questions would be the first database tool in the area of string solving which allows the extraction of information from string solving benchmarks and fair and uniform comparison of string solvers on a selected set of benchmarks displaying certain particularities. 
Such a tool could also open the way towards deeper research tasks related to the evaluation of the solvers' performance, such as analysing the impact that the preprocessing part executed by a solver has on the performance, or integrating external tools in the database, allowing the generation of new instances based on existing benchmarks.
Also, such a tool would fit in the direction of creating large collections of benchmarks containing SMT or SAT instances \cite{smtbenchmarks,satbenchmarks}.

\paragraph{Our contribution.}
We propose \toolname{}: a benchmark analysis tool for string constraints. It is accessible at: 
\toolWeb{}.

\toolname{} is implemented in \textsc{Python 3}, and offers a collection of analysis and information extraction tools for the most comprehensive database of string benchmarks (collected from the literature and presented in SMT-LIB format), based on an SQL-centred language called \qlang{}. Besides basic database management, benchmark querying, and analysis capabilities, \toolname{} also offers an interface to running and testing string solvers on the benchmarks. The results of such runs can then be collected, stored, further analysed, and correlated to other properties of the respective benchmarks (computed using \toolname{} database queries). As such, \toolname{} offers solutions to our two research tasks. \toolname{} also offers users a simple method for implementing and running their own analysis on the benchmarks, as well as the possibility of collecting and integrating the results of this analysis into the database. The user base, architectural details, use cases of \toolname{} are discussed in the rest of the paper and in the documentation of the tool.

Worth mentioning already at this point: our tool is focused, so far, on benchmarks (and underlying theories) related to string solving. This tool can be extended to cover other theories, as well, and, as such, offers the foundations for creating a more comprehensive data base for SMT-solving in general. 

\section{Potential Applications of \toolname{} and User Base} 
We begin with examples from the literature, where this tool could have been used. In particular, our goal in this section is to show that there is a demonstrable need for analyses of the properties of benchmarks containing string constraints. In the following, we overview several cases where hand-crafted benchmarks and ad-hoc analyses were created and used. \toolname{} offers a more general and easier-to-use framework for such analyses.

The first example for an ad-hoc analysis is \cite{ChenCHLW18}, which motivates the Straight-line fragment of string constraints by noting (as the product of the aforementioned analysis) that the Kaluza benchmark set falls within it. As the Kaluza benchmarks become older and new sets emerge, it would be beneficial to have similar updated analyses for these new sets of constraints, so that assumptions about practical instances do not become out-of-date. \toolname{} is aimed to make these much quicker and easier. \looseness=-1

In \cite{LeH18} the authors argue that a new, hand-crafted benchmark, whose instances fulfil some specific properties, is required to be able to compare their decision procedure with existing solvers. This paper could have benefited from our tool. In such cases, it would be faster and more transparent to use \toolname{} to extract from existing benchmarks the instances exhibiting the addressed properties, and showcase the respective novel algorithms on already existing, provably relevant instances. \looseness=-1

A similar argument can be made related to \cite{Chen2021}, where the authors say ``\emph{There are no standard string benchmarks involving RegExes[...]}". Using \toolname{} such benchmarks could be extracted from existing instances.

Nevertheless, in \cite{cav21}, a set of over 100000 benchmarks was analysed ad-hoc, to extract string constraints containing only regular expressions and linear arithmetic and detect their structural complexity, with the ultimate goal of producing an efficient solver for such constraints. This required a complex and tedious approach. Using \toolname{}'s abilities potentially simplifies this process.

Therefore, it seems that analyses like those from the aforementioned examples 
require significant effort, replicated every time a new analysis is needed. We expect \toolname{} to be used routinely by developers of string-solvers (like those mentioned in, e.g., \cite{Amadini2020,Hague19}) to provide faster and transparent evaluation, up-to-date context, and applicability evidence for their algorithms.\looseness=-1

Similarly, many theoreticians use string-solving as motivation for their works. \toolname{} can provide real evidence supporting this and also inform future directions of study. To this end, Hague~\cite{Hague19} presents examples of research groups likely to use \toolname{}.\looseness=-1

\medskip

In conclusion, the cases overviewed above, as well as examples found in literature, immediately reveal three different communities of potential users of \toolname{}: firstly, the community of string solver developers, for which it eases the performance-analysis of specific solvers, on specific benchmarks, and, thus, helps in discovering the strengths and weaknesses of each string solver, for instance by identifying certain features of the input and correlating them to the solver's performance. Secondly, the theory community: \toolname{} facilitates the further understanding of structural properties of specific classes of word equations, relevant in practice, which can be the focus of theoretical investigations. Finally, the end-users, not explicitly mentioned before: entities who have (or develop) use cases with string constraints, of relevance to their activities, and want to understand better the nature of these benchmarks w.r.t. standard structural measures for string constraints, or solve their instances as correctly and efficiently as possible; for them, producing their own analysis tools or solvers could be too expensive so they could integrate their cases in \toolname{}, and use the offered methods to analyse it.\looseness=-1

\section{Architecture of \toolname{}}

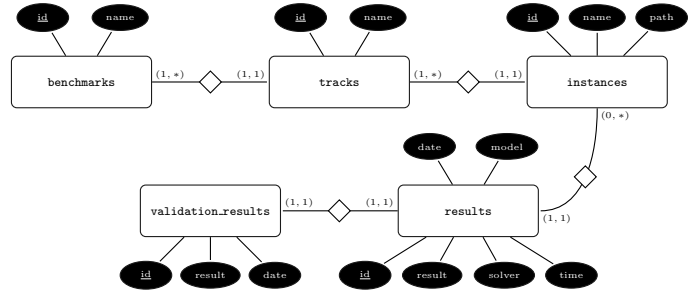
\begin{figure}[t!]
\centering
\resizebox{.75\textwidth}{!}{
 \begin{tikzpicture}[-,>=stealth',shorten >=1pt,auto,node distance=5cm,
        semithick]
        \tikzset{t/.style={draw=black!90,fill=none,text=black, text width=2.5cm, align=center,font=\scriptsize\ttfamily,rounded corners=3pt,minimum height=1cm,fill=none},
                 n/.style={draw=black!90,fill=black,text=white, align=center,font=\tiny,rounded corners=3pt,ellipse,minimum width=1cm,minimum height=0.5cm}}

        \node[t] (benchmark)   {benchmarks};
        \node[n] (benchmark_id) at ($(benchmark)+(-0.75cm,1.25cm)$)  {\underline{id}};
        \node[n] (benchmark_name) at ($(benchmark)+(0.75cm,1.25cm)$)  {name};
        \path (benchmark) edge [] node[] {}  (benchmark_id)
                          edge [] node[] {}  (benchmark_name);

        \node[t,right of=benchmark] (track)       {tracks};
        \node[n] (track_id) at ($(track)+(-0.75cm,1.25cm)$)  {\underline{id}};
        \node[n] (track_name) at ($(track)+(0.75cm,1.25cm)$)  {name};
        \path (track) edge [] node[] {}  (track_id)
                          edge [] node[] {}  (track_name);

        \node[t,right of=track] (instance)       {instances};
        \node[n] (instance_id) at ($(instance)+(-1.25cm,1.25cm)$)  {\underline{id}};
        \node[n] (instance_name) at ($(instance)+(0cm,1.25cm)$)  {name};
        \node[n] (instance_path) at ($(instance)+(1.25cm,1.25cm)$)  {path};
        \path (instance) edge [] node[] {}  (instance_id)
                          edge [] node[] {}  (instance_path)
                          edge [] node[] {}  (instance_name);

        \node[t] (result) at ($(instance)+(-2.5cm,-2.5cm)$)       {results};
        \node[n] (result_id) at ($(result)+(-2cm,-1.25cm)$)  {\underline{id}};
        \node[n] (result_result) at ($(result)+(-0.7cm,-1.25cm)$)  {result};
        \node[n] (result_solver) at ($(result)+(0.7cm,-1.25cm)$)  {solver};
        \node[n] (result_time) at ($(result)+(2cm,-1.25cm)$)  {time};
        \node[n] (result_model) at ($(result)+(0.75cm,1.25cm)$)  {model};
        \node[n] (result_date) at ($(result)+(-0.75cm,1.25cm)$)  {date};

        \path (result) edge [] node[] {}  (result_id)
                          edge [] node[] {}  (result_result)
                          edge [] node[] {}  (result_date)
                          edge [] node[] {}  (result_model)
                          edge [] node[] {}  (result_time)
                          edge [] node[] {}  (result_solver);

        \node[t, left of=result] (validation)       {validation\_results};
        \node[n] (validation_id) at ($(validation)+(-1.25cm,-1.25cm)$)  {\underline{id}};
        \node[n] (validation_name) at ($(validation)+(0cm,-1.25cm)$)  {result};
        \node[n] (validation_path) at ($(validation)+(1.25cm,-1.25cm)$)  {date};
        \path (validation) edge [] node[] {}  (validation_id)
                          edge [] node[] {}  (validation_path)
                          edge [] node[] {}  (validation_name);

        \path (benchmark) edge [] node[diamond,draw=black,fill=white,anchor=center] {}  (track)
              (track) edge [] node[diamond,draw=black,fill=white,anchor=center] {}  (instance)
              (instance) edge [out=270,in=0] node[diamond,draw=black,fill=white,anchor=center] {}  (result)
              (result) edge [] node[diamond,draw=black,fill=white,anchor=center] {}  (validation);

        \node[font=\tiny] at ($(benchmark.east)+(0.35cm,0.15cm)$)  {$(1,*)$};
        \node[font=\tiny] at ($(track.west)+(-0.35cm,0.15cm)$)  {$(1,1)$};
        \node[font=\tiny] at ($(track.east)+(0.35cm,0.15cm)$)  {$(1,*)$};
        \node[font=\tiny] at ($(instance.west)+(-0.35cm,0.15cm)$)  {$(1,1)$};
        \node[font=\tiny] at ($(instance.south)+(0.35cm,-0.15cm)$)  {$(0,*)$};
        \node[font=\tiny] at ($(result.east)+(0.35cm,-0.15cm)$)  {$(1,1)$};
        \node[font=\tiny] at ($(validation.east)+(0.35cm,0.15cm)$)  {$(1,1)$};
        \node[font=\tiny] at ($(result.west)+(-0.35cm,0.15cm)$)  {$(1,1)$};

                \end{tikzpicture}
}
\caption{\toolname{}'s \textsc{SQLite} database schema}
\label{fig:sqliteDB}
\end{figure}

We begin the technical part of this paper by discussing the main ideas behind the architecture of \toolname{}.

\toolname{} provides a series of mechanisms easing the access to a comprehensive set of benchmarks, based on an \textsc{SQL}-inspired query language called \qlang{}. The tool is built such that it can be run on an everyday workstation within a terminal and it aims to provide answers to the user's questions regardless of the time it takes to get them. To this extent, we have tried to make \toolname{} as flexible as possible, giving easy-to-use entry-points to adding custom algorithms for the analysis of string solvers or benchmarks without requiring high-performance servers. Nevertheless, due to multiprocessing, we allow running our tool in a server environment which speeds up the answering of the asked questions, providing rather superb response times, as we discuss in Section~\ref{sec:examplesUSE}. The input is given in our novel query language called \qlang{}, which allows accessing and analysing benchmarks following the SMT-LIB standard regardless of their origin, directly in a terminal prompt. To implement the main structure of our database of string constraints-benchmarks, we proceeded as follows. The central information is stored in an \textsc{SQLite} database which consists of five different tables, namely \texttt{benchmarks}, \texttt{tracks}, \texttt{instances}, \texttt{results}, and \texttt{validation\_results} visualised in Figure~\ref{fig:sqliteDB}. Each benchmark set contains multiple tracks (e.g., \textsc{Kaluza} contains different tracks, grouping instances w.r.t. their size and satisfiability), which is reflected in our database structure via the tables \texttt{benchmarks} and \texttt{tracks}. A track itself contains multiple linked instances, stored within the \texttt{instances} table. Thus, the \texttt{instances}-table stores for each record a file path and additional information, such as a unique name. Initially, a new benchmark set including its instances is always registered in our database. Since we also provide an interface for running different solvers on the available benchmarks, we allow storing the results of these runs in the database, in the \texttt{results} table. These results are cross-validated w.r.t. the existing solvers and the conclusions are stored in table \texttt{validation\_results}. 

We provide an easy interface, allowing us to add additional solvers. Currently, \toolname{} implements it for \cvc{}, \textsc{Z3Str3}, and \textsc{Z3Seq}, but can be extended to include string solvers capable of reading/processing SMT-LIB files. To achieve this, we reused the engine of \textsc{ZaligVinder}. We took the scheduling engine allowing multiprocess runs of solvers, as well as the runners developed for the benchmark framework. This includes special handling for different string-solvers as explained in \cite{zaligVinderJournal}. Furthermore, we reuse the cross-validation mechanism: \textsc{ZaligVinder} runs all competing solvers and whenever a server returns \texttt{SAT}, we check the validity by asserting the model into the original instance and using another solver to check correctness. In the case of \texttt{UNSAT} and when no other solver returned a valid model, we use a majority vote upon all solvers' results.\looseness=-1 

To allow gathering new insights about the benchmarks, \toolname{} offers an interface permitting the definition of custom benchmark-analysis predicates, which can directly interact with the SMT-LIB instances and the pre-calculated information regarding them. To this end, for each instance contained in the database, we additionally store an Abstract Syntax Tree (AST) within the file system and have the possibility to augment each node of the tree with additional information. These ASTs are the fundamental data structures used in our approach.\looseness=-1 

\paragraph{The language \qlang{}.} Let us now go a bit more into details regarding the query language \qlang{}. As its main functionality, this language allows the selection of instances (i.e., printing file names matching a query or exporting them after potentially applying a modification). The syntax of \qlang{} is given in Figure~\ref{fig:qlang}.

\begin{figure}[t!]
\fontsize{8}{8}\selectfont
\[
\begin{array}{lcll}
  S & \Coloneqq & \texttt{Select } f_s \texttt{ From } d \texttt{ Where } c \!\vsep\! \texttt{Extract } f_e \texttt{ From } d \texttt{ Where } c \texttt{ Apply } \textsc{Function} \\
  f_s & \Coloneqq & \texttt{Name} \vsep \texttt{Hash} \vsep \texttt{Content}\\
  f_e & \Coloneqq & \texttt{SMTLib} \vsep{} \texttt{SMTPlot}
                    \vsep \dots\\ 
  d & \Coloneqq & \texttt{*} \vsep \textsc{Set} \vsep \textsc{Set}\texttt{:}\textsc{Track} \vsep d\texttt{, } d\\
  c & \Coloneqq & \textsc{Predicate} \vsep (c \texttt{ And } c) \vsep (c \texttt{ Or } c)  \vsep (\texttt{Not } c) \vsep \texttt{True} \vsep \texttt{False}\\
  \end{array}
\]

\caption{Syntax of \qlang{}.}
\label{fig:qlang}
\end{figure}
The semantic of a \qline{Select} query is based on the data set $d$. We either choose all benchmarks (\qline{*}) or we are more precise in picking a particular benchmark set respectively a corresponding track.

\begin{figure}[b!]
\centering
\resizebox{0.75\textwidth}{!}{
 \begin{tikzpicture}[->,>=stealth',shorten >=0pt,auto,node distance=3cm,
        semithick]
        \tikzset{t/.style={color=white,fill=black,circle,minimum width=0.1cm,font=\tiny},
                 n/.style={draw=black!90,fill=none,text=black, align=center,font=\tiny,rounded corners=3pt,ellipse,minimum width=1cm,minimum height=0.5cm},
                 box/.style={draw=black!90,minimum width=0.25cm,minimum height=0.25cm,node distance=0.25cm,rounded corners=1pt}}

        \newcommand{\getIntel}[2]{
                \node [right of=#1,box,node distance=0.65cm,#2] (#10) {};
                \foreach \x [count=\i] in {0,1,...,5} {
                  \node [box,right of=#1\x,#2] (#1\i) {};
                } 
        }

        \node[t] (A)   {A};
        \getIntel{A}{solid}
        \node[t,below left of=A] (B)   {B};
        \getIntel{B}{solid}
        \node[t,below right of=A] (C)   {C};
        \getIntel{C}{solid}

        \node[node distance=1cm,above of=A] (top)   {$\vdots$};

        \path (A) edge [out=250,in=90] node[] {}  (B)
                  edge [out=290,in=90] node[] {}  (C)
                  edge [<-] node[] {} (top);

        \draw [decorate, decoration = {calligraphic brace,mirror},-] ($(B0.south west)+(0,-0.1cm)$) --  ($(B6.south east)+(0,-0.1cm)$);
        \node[] (B_f) at ($(B3.south)+(0,-0.5cm)$)   {$f(B,\varepsilon)$};

        \draw [decorate, decoration = {calligraphic brace,mirror},-] ($(C0.south west)+(0,-0.1cm)$) --  ($(C6.south east)+(0,-0.1cm)$);
        \node[] (C_f) at ($(C3.south)+(0,-0.5cm)$)   {$f(C,\varepsilon)$};

        \node[] (M) at ($(A)+(2cm,-1cm)$)  {};
        \getIntel{M}{solid,black!50}
        \path (M0) edge [out=180,in=90,<-,dashed,black!50,shorten <=1pt] node[] {}  (B3)
              (M5) edge [out=270,in=90,<-,dashed,black!50,shorten <=1pt] node[] {}  (C3);
        \node[anchor=west,black!50] (M_m) at ($(M6.east)+(0,0cm)$)   {merge of intels $B$ and $C$};

        \draw [decorate, decoration = {calligraphic brace},-] ($(A0.north west)+(0,0.1cm)$) --  ($(A6.north east)+(0,0.1cm)$);
        \node[] (A_f) at ($(A3.north)+(0.25cm,0.5cm)$,anchor=west,align=left)   {$f(A,\hspace{1.8cm})$};
        \node[] (AM) at ($(A3.north)+(-0.9cm,0.5cm)$)  {};
        \getIntel{AM}{solid,black!50}
        \path (AM5) edge [out=270,in=90,<-,dashed,black!50,shorten <=1pt] node[] {}  (M3);

                \end{tikzpicture}
}
\caption{Calculation of the \textsc{IntelDictionary} in our AST. $f$ is the \texttt{apply}-function, $\varepsilon$ the corresponding neutral element, and $A,B,C$ arbitrary nodes and their \textsc{IntelDictionary} (boxes next to nodes) in our AST corresponding to an SMT-LIB instruction.}
\label{fig:ASTstructure}
\end{figure}
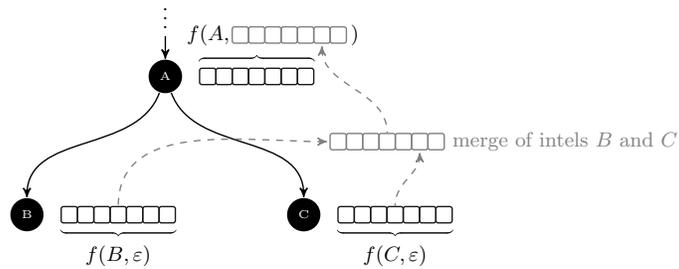

The selection is based on a Boolean expression $c$, constructed from basic \textsc{Predicate}s. Currently, \toolname{} implements several default predicates. For instance, the default predicate \qline{hasWEQ} selects the benchmarks containing at least one word equation, or \qline{isSAT($solver$)} returns instances being declared satisfiable by the particular $solver$. Worth noting, our interface allows also defining custom predicates. As far as the implementation is concerned, when evaluating a predicate (as, for example, our default predicates) on an instance, this predicate is applied bottom-up to the corresponding AST and its value is evaluated in the root. Therefore, in the case of custom predicates, the user has to specify (just as we did for the default predicates) two functions, namely an \texttt{apply}- and a \texttt{merge}-function. The \texttt{apply}-function specifies how the actual computation of the predicate is done for the information corresponding to a single node, while the \texttt{merge}-function processes the data computed by the children of the argument node. The related information attached to each node is called \textsc{IntelDictionary} which will be explained in detail within the next paragraph. We visualise this procedure in Figure~\ref{fig:ASTstructure}. As a basic, yet illustrative example, consider the word equation $a\mathsf{Y}ab\mathsf{X} \doteq \mathsf{Z}abb\mathsf{Y}$ where $a$ and $b$ are terminals and $\mathsf{X}, \mathsf{Y}$ and $\mathsf{Z}$ are variables. The calculation of the number of occurrences of each variable in a node starts by counting these occurrences within the two sides of the equation leading to the sets $\Set{(\mathsf{X},1),(\mathsf{Y},1)}$ and  $\Set{(\mathsf{Y},1),(\mathsf{Z},1)}$. Our root node corresponds to the equality $\doteq$. We apply the merge-function, which adds up the occurrences of variables in the children nodes of the current node, to obtain the set $\Set{(\mathsf{X},1),(\mathsf{Y},2),(\mathsf{Z},1)}$ which indeed is the desired data for the root node, since $\doteq$ does not contain any other variables. In general, the actual predicate uses the computed data for an AST and returns either \texttt{true} or \texttt{false}, thus allowing the selection of particular instances based on this return value. Continuing our previous example, asking whether a word equation is quadratic via the predicate \qline{isQuadratic} can simply use the previously calculated data (number of occurrences of the variables) and simply return \texttt{true} if and only if each variable has at most two occurrences. 

Finally, we use $f_s$ to choose a suitable output which can be the instance name, the file's hash value, or simply the SMT-LIB instance.

%
%

The \qline{Extract} query allows exporting instances for which a Boolean expression (again involving predicates) evaluates to \texttt{true}, just as described above for \qline{Select} queries. Additionally, while executing a \qline{Extract} query, we can directly perform modifications to the extracted instances using a \texttt{Function}. These functions are applied (similarly to the case of predicates) node-wise, bottom-up, on the ASTs corresponding to the processed instances. Therefore, for such queries, the user specifies for the nodes an \texttt{apply}-function, which performs the modification of a node. This technique allows, for example, applying simplification rules to specific nodes or simply restricting an instance to a particular kind of string constraint (e.g. word equations). Moreover, this interface allows the application of external procedures on the whole AST, such as applying a fuzzer like \textsc{StringFuzz}~\cite{blotsky2018stringfuzz} to generate new instances having a similar structure to the extracted ones. Finally, the argument $f_e$ of an \qline{Extract} query specifies the output format for the matching (and potentially modified) instances, e.g. in SMT-LIB format or a plot visualizing the tree-like structure. As a simple example, to count instances which exhibit certain properties, one could use the predefined operation \qline{Count}. 
Notably, a function might also translate ASTs into different (not necessarily tree-like) structures, providing, in a sense, an interpretation of these trees suitable to the desired application.\looseness=-1

As an example for a query, to obtain a list of all benchmarks containing word equations and determined satisfiable by the string solver \cvc{}, we can execute the query \qline{Select Name From * Where (hasWEQ and isSAT(CVC5))}. A second example is removing all other constraints than word equations from our benchmarks and exporting the resulting SMT-LIB files. We use an \qline{Extract} query and pose \qline{Extract SMTLib From * Where hasWEQ Apply Restrict2WEQ}\footnote{A list of the available options is printed when executing our tool and available in \toolname{}'s documentation}.


\paragraph{The AST structure.} At the core of our implementation is the AST data structure, which is directly derived from the SMT-LIB instances. A string constraint defined in SMT-LIB contains variable declarations and (potentially) multiple asserts of formulae being based on string constraints connected by the common connectives (note that the string constraints are not quantified in the respective standard). All asserted formulae have to be satisfied at the same time. Based on this structure, our AST is a parse tree derived according to the formal grammar from Figure~\ref{fig:ast}, modelling each formula. As it can be seen there, \toolname{} currently supports common string-constraints: word equations, linear arithmetic-over-lengths, regular-language\--mem\-ber\-ship, Boolean constraints. We use \textsc{Z3} as input parser for SMT-LIB, so \toolname{} parses all constraints \textsc{Z3} handles. Our internal representation uses a generic expression to represent constraints/types not covered explicitly yet in our grammar. Thus, as already hinted at the end of the Introduction, \toolname{} can be canonically extended to address other constraint types and, as such, other theories. \looseness=-1
\begin{figure}[t!]
\fontsize{8}{8}\selectfont
\[
\begin{array}{lcll}
  Expr & \Coloneqq & \texttt{(\,} Id\texttt{, }Kind\texttt{, }Decl\texttt{, }Sort\texttt{, }Params\texttt{, }Children\texttt{, }\textsc{IntelDictionary} \texttt{\,)}\\
  Id & \Coloneqq & x \hspace*{0.5cm}\text{\scriptsize for $x \in \mathds{N}$} \hspace*{1.5cm} Kind \Coloneqq \texttt{Variable} \vsep \texttt{Other}\\
  Decl & \Coloneqq & \texttt{and} \vsep \texttt{or} \vsep \texttt{not} \vsep \dots \vsep \texttt{=} \vsep \texttt{<=} \vsep \dots \vsep \texttt{substr} \\
  Sort & \Coloneqq & \texttt{String} \vsep \texttt{Bool} \vsep \texttt{RE} \vsep \texttt{Integer}\\
  Params & \Coloneqq & \texttt{[} p \texttt{]} \vsep \texttt{[]} \hspace*{2cm} p  \Coloneqq  x \vsep p\texttt{, }p\hspace*{0.5cm}\text{\scriptsize for $x \in \mathds{Z}$}\\ 
  Children & \Coloneqq & \texttt{[} c \texttt{]} \vsep \texttt{[]} \hspace*{2cm} c  \Coloneqq  Expr \vsep c\texttt{, }c\\

  \end{array}
\]
\caption{Internal representation of string constraints.
}
\label{fig:ast}
\end{figure}

Let us now go into some of the technical details on which the parsing process and the ASTs are based. We begin by explaining the grammar of Figure~\ref{fig:ast}, which lays the foundations for the internal representation of string constraints. In this grammar, each expression $Expr$ (which  corresponds to a SMT-LIB formula) has a unique $Id$, a named operator which can essentially be any operator available in the SMT-LIB for string constraints, a $Kind$ declaring whether the given expression is a single variable or not, and a $Sort$. Additionally, an expression might have additional parameters; for instance, \texttt{re.loop} which corresponds to a bounded Kleene star operation w.r.t. the parameters. Furthermore, an expression can have multiple children which are again expressions. Finally, each expression stores a unique \textsc{IntelDictionary} containing all the information computed using the previously introduced predicates and stored in the database. This structure allows accessing individual nodes quickly. To avoid recalculating the ASTs over and over again, whenever needed, we use \textsc{Pickle}~\cite{pickle} to store the tree within the file system. As such, an AST corresponding to a particular instance (file) is available and can be enriched at any time. This allows quickly re-accessing of stored information, since, e.g. for selecting all instances containing word equations, only the root node's \textsc{IntelDictionary} has to be checked when using the predicate \qline{hasWEQ}.\looseness=-1

One key aspect of \toolname's architecture is the usage of the ASTs in the definition of predicates, functions, and extractors. While defining meaningful and efficient predicates/functions is an algorithmic problem, which needs to be addressed individually for each predicate/function, our architecture offers both a fundamental data structure, easily and naturally adaptable to specific scenarios, as well as an accessible interface for defining and implementing those functions.\looseness=-1 

\begin{figure}[b!]
\resizebox{\textwidth}{!}{
 \begin{tikzpicture}[->,>=stealth',shorten >=1pt,auto,node distance=4cm,
        semithick]
        \tikzset{t/.style={draw=black!90,fill=none,text=black, text width=2cm, align=center,font=\small\sffamily,rounded corners=3pt,minimum height=1cm},
                 nofill/.style={draw=none,fill=none}}
        \node[] (0)   {\qlang{} query};
        \node[] (1) at ($(0)+(0,-1cm)$)  {Result};
        \node[t,nofill,circle,draw=black,text width=1cm] (logic) at ($(0)+(4,-0.5cm)$)  {\faCogs\\ Logic};
        \node[t,nofill] at ($(logic)+(-1.5cm,2.5cm)$) (DB)   {\faDatabase\\Database};
        \node[t,nofill] at ($(logic)+(1.5cm,2.5cm)$) (solvers)   {\faRecycle\\Solvers};
        \node[t] at ($(logic)+(3.5cm,1.2cm)$) (pred)   {Predicates};
        \node[t] at ($(logic)+(3.5cm,0cm)$) (extr)   {Extractors};
        \node[t] at ($(logic)+(3.5cm,-1.2cm)$) (fun)   {Functions};

         \node[t,nofill,text width=5cm] at ($(extr)+(4cm,1cm)$) (fs)   {\normalfont Filesystem};
        \node[t,nofill,text width=5cm] at ($(extr)+(4cm,-0.75cm)$) (smt)   {\faFile\\\scriptsize SMT-LIB Instance};
        \node[t,nofill] at ($(extr)+(4cm,0.25cm)$) (ast)   {\faTree$^\ast$\\\scriptsize AST w/ intel};


        \path (pred) edge [out=0,in=180] node[] {}  ($(ast.north west)+(-0.45,-0.25)$)
              (fun) edge [out=0,in=180] node[] {}   ($(ast.south west)+(-0.45,-0.25)$)
              (extr) edge [out=0,in=180] node[] {}  ($(ast.west)+(-0.45,-0.25)$)
              (0) edge [dashed,out=0,in=170] node[] {}  (logic)
              (logic) edge [dashed,in=0,out=190] node[] {}  (1)
              (logic) edge [] node[font=\tiny] {run solvers}  (solvers)
              (solvers) edge [] node[font=\tiny] {store results}  (DB)
              (logic) edge [<->] node[font=\tiny,text width=0.5cm] {obtain data}  (DB)
              (logic) edge [in=180,out=20] node[] {}  (pred)
              (logic) edge [in=180,out=0] node[] {}  (extr)
              (logic) edge [in=180,out=-20] node[] {}  (fun)
              ; 

        \draw [<->] (logic) -- ($(logic)+(0,-2cm)$) -- node[font=\tiny,yshift=-0.5cm] {access files, write ASTs and extractor output} ($(logic)+(7.5cm,-2cm)$) -- ($(logic)+(7.5cm,-1.22cm)$);
        \draw [] (solvers) -- node[font=\tiny,yshift=-0.35cm] {access files} ($(solvers)+(6cm,0cm)$) -- ($(solvers)+(6cm,-1.22cm)$);
        \draw [] ($(fs.north east)+(-1cm,-.2cm)$) -- ($(fs.north west)+(1cm,-.2cm)$) -- ($(smt.south west)+(1cm,0)$) -- ($(smt.south east)+(-1cm,0)$) -- cycle;

                \end{tikzpicture}
}
\caption{Architectural overview of \toolname{}.}
\label{fig:smtquery}
\end{figure}
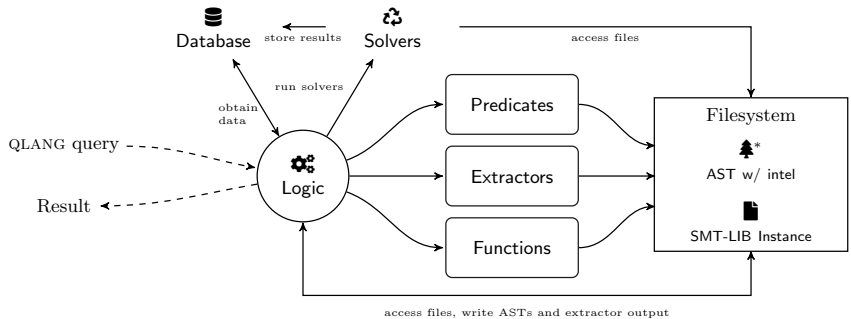

\paragraph{Summary.} In Figure~\ref{fig:smtquery} we overview the overall architecture of \toolname{}. A user poses a \qlang{} query $q$ to the command line interface. After parsing the query $q$ the core logic acquires relevant information about the selected benchmarks and schedules solver runs at any time needed. The query $q$ either has the form $\text{\qline{Select } } f_s \text{ \qline{ From } } d \text{ \qline{ Where } } c$ or  $\text{\qline{Extract } } f_e \text{ \qline{ From } } d \text{ \qline{ Where } } c \text{ \qline{ Apply } } f$ as opposed in the grammar shown in Figure~\ref{fig:qlang}. For the selection of the benchmarks $d$, we query our \textsc{SQLite} database which returns related information (i.e. a pointer to the AST, a unique id, and file-system path) for each requested instance. We apply the predicate $c$ to each instance obtained from the previous query, potentially removing it from our selection. The predicate $c$ makes use of the \textsc{IntelDictionary} stored in our ASTs. When the AST is not available, the \zthree{}'s output of the parsed SMT-File is translated into our AST. Furthermore, if parts of the requested data in the \textsc{IntelDictionary} are not available, we recalculate it on the fly and additionally enrich our AST with the newly obtained information. We do not store the \textsc{IntelDictionary} within our \textsc{SQLite} database to stay as flexible as possible. Since each node of our AST enriches the \textsc{IntelDictionary} of its children, storing this information inside our database would result in storing a link to each node. Secondly, adding new entries to the \textsc{IntelDictionary} would require a modification of our database schema. A predicate $c$ might also ask for solver related information (e.g. \qline{isSAT(Z3Str3)}, asking for all instances declared satisfiable by \zstr{}). In this case, we again query our \textsc{SQLite} database for the requested information. Whenever the data is not available, we automatically call the solver and store the corresponding results within the database, making it accessible for further queries. The same step calls the verification mechanism explained earlier. Posing a \qline{Select} query to \toolname{} outputs the results being specified by $f_s$ directly into the user's terminal. An \qline{Extract} query acts differently depending on the extractor $f_e$. As explained previously, an extractor can modify matching instances based on its own needs. Therefore, $f_e$ might write data to the file-system (e.g. a cactus plot using \qline{CactusPlot} or an SMT-File using \qline{SMTLib}) or simply print a result to the user's terminal (e.g. a summary of solver results using \qline{InstanceTable}). If the query contains a function $f$ within it \qline{Apply}-part, the core logic performs modifications according to the specification given by $f$ before using the extractor $f_e$. 

\smallskip

The operations implemented so far in \toolname{} heavily differ in their run time. Clearly, it is inherent that some operations require a rather long execution time: firstly, we analyse a huge set of instances, and, secondly, the analysis we apply might involve complicated predicates, which are provably computationally hard. Our approach to speeding up this process is to allow the incremental inclusion in the stored data of the results of various queries, on which one can build efficiently more complex queries. Summing up, our goal was to build a tool allowing information extraction from benchmarks of string constraints, using a state of the art home-computer, without having to go deep into implementation details. In the next second, we report the running times of executed queries.

\medskip

Further details are given in \toolname{}'s online documentation at \toolWeb{}.\footnote{To ease the reviewing process, some examples are given in the Appendix.}

\section{Use cases and examples}
\label{sec:examplesUSE}
This section is devoted to examples of problems which can be addressed with \toolname{}. We pose a task and describe the difficulties arising while solving it. Afterwards, we show how certain problems can be addressed using \toolname{} and support our approach with results and statics based on \textsc{ZaligVinder}'s benchmark set.

\medskip

\begin{table}[t!]\centering \resizebox{.5\textwidth}{!}{\begin{tabular}{*{3}{c}}
	SMT-LIB 2.5 keyword & ~~ &SMT-LIB 2.6 keyword \\
	\hline
	\texttt{int.to.str} &  &\texttt{str.from\_int}\\
    \texttt{str.to.int} &  &\texttt{str.to\_int}\\
    \texttt{str.in.re}  &  &\texttt{str.in\_re}\\
    \texttt{str.to.re}  &  &\texttt{str.to\_re}\\
    \texttt{re.nostr}   &  &\texttt{re.none}\\
    \texttt{re.empty}   &  &\texttt{re.none}\\
    \texttt{\textbackslash x0$n$} & & \textbackslash u\{$n$\}\\
    \texttt{\textbackslash x$m$} & & \textbackslash u\{$m$\}
\end{tabular}}
\vspace*{0.2cm}

\caption{Translation from SMT-LIB 2.5 to 2.6 for $n \in \mathds{N}_{\leq 9}$ and $m \in \mathds{N}_{>9}$}
\label{tab:smttrans}

\vspace*{-0.75cm}
\end{table}

Our experimental setup is built upon $114468$ different SMT-LIB instances gathered in \cite{zaligVinderJournal} containing 19 different sets mainly stemming from real-world applications and solver developers as explained in our introduction. Firstly, to incorporate the most recent release of \cvc{}, we manually translated these benchmarks into SMT-LIB 2.6. The gathered instances were still in SMT-LIB 2.5 format which is no longer supported by \cvc{}. The translation itself is a straightforward renaming of the keywords and functions given in Table~\ref{tab:smttrans}. We set up \toolname{} on a server running Ubuntu 18.04.4 LTS with two AMD EPYC 7742 processors having a total of 128 cores and 2TB of memory. We integrated \cvc{}'s version 1.0.1 and \textsc{Z3}'s version 4.10.1 binaries from their official sources.

Before we consider actual use cases, we use \toolname{} to get a better intuition on the used benchmarks and obtain some insights. First, we initialise the \textsc{SQLite} database such that it contains the schema shown in Figure~\ref{fig:sqliteDB} and links all of our $114468$ instances accordingly. This process took $4.44$ minutes. We now use our built-in predicates to observe that $80284$ instances contain word equations (using the predicate \qline{hasWEQ}), $57257$ contain regular-expression membership constraints, and $59763$ contain linear arithmetic over string length. Additionally, we discovered that $30393$ contain higher-order functions (e.g. \texttt{str.substring}, \texttt{str.replace}). The running time for each query without using the cache was about $7.16$ minutes. Using our pre-cached ASTs allows us to acquire the above values in roughly $70$ seconds.

The above values do not require satisfiability results of the embedded string solvers. As mentioned in the previous section, all integrated solvers will be run automatically. To quickly access cached results, our tool allows running all solvers (including verification) in advance. The running times heavily depend on selected timeouts and machine power, as well as the performance of the embedded solvers (e.g. obtaining results takes longer if a solver times out more often). To give an intuition on the running times using previously stored results, we discover that \cvc{} declares $71540$ instances satisfiable. We obtain this value in 5.13 minutes.

\medskip

We now move on to particular use cases. The first generic problem we address is the following:

\begin{quote}
	\begin{description}
	\item {\sffamily Problem I:} \emph{Given a syntactically restricted subset of string constraints, determine instances belonging to this subset, and their distribution w.r.t. benchmarks.} 
	\end{description}
\end{quote}
Many theory papers provide insightful results (i.e., algorithms, complexity bounds, information about solution sets) for subclasses of string constraints \cite{Hague19}. Such subclasses include those defined directly (e.g., constraints in solved-form, acyclic or straight-line constraints) as well as indirectly (e.g., quadratic or regular word equations). Knowing how applicable such results resp. insights are in practice, and thus whether it makes sense to incorporate them in the design and implementation of string solvers, requires first knowing how many string constraints belong to those subclasses. By solving this, \toolname{} is valuable for researchers approaching subcases of string constraints, who could use our tool to see if that subcase is relevant to string-constraint solving in practice, and if so, provide evidence of this as motivation for their work. If the defined subcase is not prominent, they could use it to guide changes to the definition, in order to make it more applicable while preserving theoretical properties. This approach is also of value to researchers developing string solvers who will benefit from knowing which theoretical insights are most likely to be effective over a broad range of use-cases and which properties to target with their own optimizations and innovations. 

When dealing with the aforementioned problem it is worth noting that, firstly, syntactic restrictions continually arise from a variety of (theory-)sources and will not necessarily be formulated directly in the usual nomenclature of string constraints and SMT-solvers. Consequently, simply deciding whether a string constraint belongs to a subclass of interest can range from trivial to requiring substantial processing. For example, it is not immediately clear given a single instance, whether e.g. the systems of word equations arising when solving it are all quadratic. Secondly, the set of benchmarks is also regularly being expanded and updated (and some might also depreciate). Thirdly, many modern string solvers rewrite string constraints in a preprocessing stage or even constantly while solving them. Obtaining realistic data, therefore, requires taking into account the effects of rewriting processes concerning syntactic subsets. 

Currently, there are no tools capable of properly addressing this problem and these challenges. Without \toolname{}, understanding e.g. how many string constraints belong to various relevant fragments is something which would have required significant effort for just a single case. In this context, it does not come as a surprise to see that such analyses have not been yet carried out even for major subclasses of string constraints.

We can give two concrete cases of the problem stated in this generic example. Firstly, we investigate the distribution of quadratic equations (a well-studied class of word equations, which can be solved using a technique inspired by Levi's lemma~\cite{choffrut1997combinatorics}) in the benchmarks. A typical analysis might result in the following questions:
\begin{enumerate}[1.]
\item For each benchmark, determine all instances consisting of quadratic equations only: \qline{Select Name From * Where isQuadratic}. In $63$ seconds using our cached AST \toolname{} outputs a list of all $47796$ matching instances.

\item Count how many such instances are in each benchmark, and compute the ratio between the number of quadratic instances and the overall number of instances in each benchmark (e.g. for JOACO-Suite): 

\begin{quote}\centering\qline{Extract Count From joacosuite Where isQuadratic}.\end{quote}

 After less than $1$ second \toolname{} reports ``\!\!\!\!\tline{Total matching instances: 51 of 94 within the selected set (54.25\%)}''.
\end{enumerate} 

Secondly, motivated by the work performed in \cite{cav21,words21}, we want to determine all instances containing regular-membership predicates, and their distribution within benchmarks.

\begin{enumerate}[1.]
\item For each benchmark, determine all instances containing at least one regular-membership predicate: \qline{Select Name From * Where hasRegex}. After roughly $70$ seconds \toolname{} prints a list of $57257$ containing regular-membership predicates.
\item Count how many such instances are in each benchmark, and compute the ratio between the number of instances containing regex-membership predicates and the overall number of instances in each benchmark (e.g. for JOACO-Suite): 

\begin{quote}\centering\qline{Extract Count From joacosuite Where hasRegex}.\end{quote} After less than $1$ second we obtain the output ``\!\!\!\!\tline{Total matching instances: 76 of 94 within the selected set (80.85\%)}''.
\item We are interested in gathering knowledge about how many of the instances containing regex-membership predicates fall into the $\mathsf{PSPACE}$-complete fragment of simple regex-membership predicates (i.e. predicates of the form $\mathsf{x}\,\dot\in\, R$, where $\mathsf{x}$ is a variable and $R$ is a regular expression not containing complements). We pose: 

\begin{quote}\centering\qline{Extract Count From * Where isSimpleRegex}.\end{quote}

 \toolname{} returns ``\!\!\!\tline{Total matching instances: 24486 of 114468 within the selected set (21.39\%).}'' in $2.10$ minutes.
\end{enumerate}

\medskip

\noindent We move on to a second generic problem.
\begin{quote}
	\begin{description}
	\item {\sffamily Problem II:} \emph{For a given string solver, understand the properties of instances on which it performs particularly well, and on which it performs poorly.} 
	\end{description}
\end{quote}

Having insights are valuable for designing new and improving or optimising existing string solvers. It is also valuable for constructing {\em portfolio solvers} who simply choose a well-performing algorithm for a particular case (see e.g. \cite{z3str4}).

Clearly, some challenges stem from the same issues discussed in the previous example. Moreover, once we computed a set of instances on which a solver performs well and a set of instances on which that solver performs poorly, we need a reliable analysis tool to find properties which separate the two sets.

Tools already exist which assist the comparison of string solvers in terms of directly evaluating how they perform over a set or sets of benchmarks (e.g., \cite{beyer2016reliable,zaligVinderJournal}). This is sufficient for producing evidence of their effectiveness within the current landscape of solvers. However, without \toolname{}, it is difficult even to understand the character of particular benchmarks beyond very superficial observations. Thus, existing tools do not provide insights about why or when a given solver performs well. Our tool is the first to facilitate analyses of the form ``\emph{Solver $X$ performs best on string constraints containing complex word equations}" where ``\emph{complex}" can be formally defined by a well-motivated criteria obtained by using \toolname{}.

We can give a concrete case related to the problem stated in this example. We would be interested in finding the set $C$ of all the instances on which \cvc{} provides a correct answer and \zstr{} either provides a wrong answer or is slower in providing the correct answer and the instances the set $Z$ of all the instances on which \zstr{} provides a correct answer and \cvc{} either provides a wrong answer or is slower in providing the correct answer. Then, for each of these sets, detect 
the number (and distribution) of instances containing regular-membership predicates. A typical analysis using \toolname{} might look as follows:

\begin{enumerate}[1.]
	\item Collect, for each instance, the answers given by all solvers included in our tool. The supposedly-correct answer for this instance is the one given by the majority of these solvers in \texttt{UNSAT} cases and indicated by a correct model in case of \texttt{SAT} instances. We pose 

	\begin{quote}\centering\qline{Extract InstanceTable From * Where ((isCorrect(CVC5) and isCorrect(Z3Str3)) and isCorrect(Z3Seq))}\end{quote}

	to \toolname{}. After $17$ minutes our terminal displays the table shown in Figure~\ref{fig:terminal}. The output might differ depending on the initialisation of the instances and scheduling of the processes. 

	\lstconsolestyle
	\begin{figure}[t!]
\centering
	\begin{lstlisting}
Waiting for results ...
Instance                 Result CVC5      Time CVC5  Result Z3Seq      Time Z3Seq  Result Z3Str3      Time Z3Str3
-----------------------  -------------  -----------  --------------  ------------  ---------------  -------------
pisa:pisa:pisa-011.smt2  Satisfied       0.00897606  Satisfied          0.0259043  Satisfied            0.0344819
pisa:pisa:pisa-009.smt2  Satisfied       0.0191097   Satisfied          0.0276228  Satisfied            0.028013
pisa:pisa:pisa-010.smt2  Satisfied       0.0167181   Satisfied          0.0258694  Satisfied            0.0266274
pisa:pisa:pisa-002.smt2  Satisfied       0.0235912   Satisfied          0.116755   Satisfied            0.0386019
pisa:pisa:pisa-000.smt2  Satisfied       0.0695572   Satisfied          0.0426866  Satisfied            0.0492182
...\end{lstlisting}
\caption{Cut terminal output for the query \texttt{Extract InstanceTable From * Where ((isCorrect(CVC5) and isCorrect(Z3Str3)) and isCorrect(Z3Seq))}}
\label{fig:terminal}

\vspace*{-0.25cm}
\end{figure}

	\item Select all instances where \cvc{} gives the right answer and either \zstr{} returns the wrong answer or it gives the right answer slower: 

	\begin{quote}\centering\qline{Select Name From * Where ((isCorrect(CVC5) and (not isCorrect(Z3Str3))) or (isCorrect(Z3Str3) and isFaster(CVC5,Z3Str3)))}.\end{quote}

	We get a list of all $94676$ matching instances within our benchmarks. This process took about $15$ minutes.
	\item Count how many of the instances computed in step 2 
	contain regular-membership predicates: 
	
	\begin{quote}\centering\qline{Extract Count From * Where (((isCorrect(CVC5) and (not isCorrect(Z3Str3))) or (isCorrect(Z3Str3) and isFaster(CVC5,Z3Str3))) and hasRegex)}.\end{quote}

	 Again, in roughly $15$ minutes we get the following answer: ``\!\!\!\!\tline{Total matching instances: 47070 of 114468 within the selected set (41.12\%)}''.
	\item Select for \zstr{} all instances where \zstr{} gives the right answer and either \cvc{} gives the wrong answer or it gives the right answer slower: 

	\begin{quote}\centering\qline{Select Name From * Where ((isCorrect(Z3Str3) and (not isCorrect(CVC5))) or (isCorrect(CVC5) and isFaster(Z3Str3,CVC5)))}.\end{quote}

	 \toolname{} prints a list of $18596$ instances within $16$ minutes.
	\item Count how many of the instances computed in step 4 
	contain regular-membership predicates: 

	\begin{quote}\centering\qline{Extract Count From * Where (((isCorrect(Z3Str3) and (not isCorrect(CVC5))) or (isCorrect(CVC5) and isFaster(Z3Str3,CVC5))) and hasRegex)}.\end{quote}

	 In $17$ minutes \toolname{} responds with ``\!\!\!\!\tline{Total matching instances: 9189 of 114468 within the selected set (8.02\%).}''
\end{enumerate}

\begin{figure}[t]
  \begin{center}
    \begin{tikzpicture}[->,>=stealth',shorten >=1pt,auto,node distance=6cm,semithick]
      \node[]   (0)  at (0,0)               {\includegraphics[width=10cm]{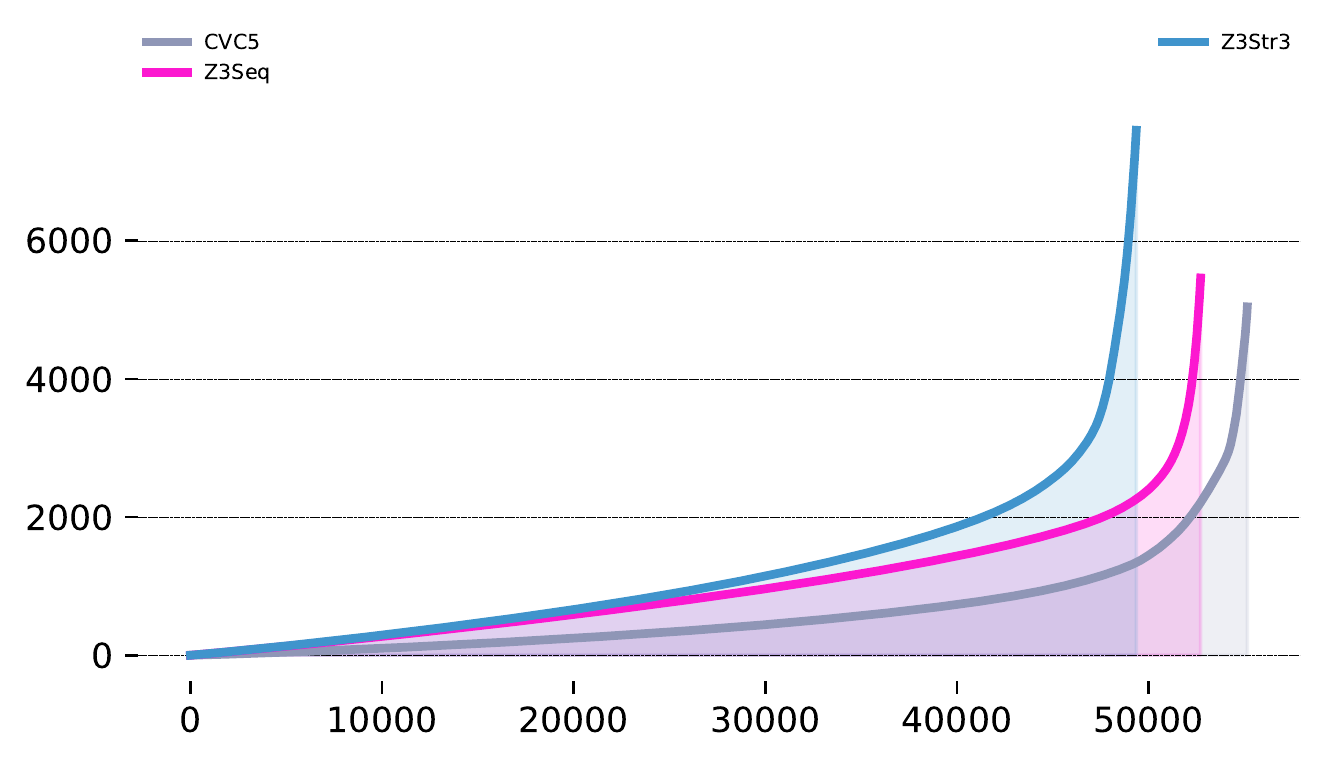}};
    \end{tikzpicture}
  \end{center}
  \vspace*{-0.5cm}
\caption{Cactus plot for query \texttt{Extract CactusPlot From * Where hasRegex}.}
\label{fig:cactusC}

\vspace*{-0.25cm}
\end{figure}

This analysis gives and substantiates an intuition that \cvc{} is more reliable than \zthree{} and seems to have a better performance when targeting instances that contain regular-membership predicates. \toolname{} offers the export of cactus plots allowing review of the results in a visually appealing way. Figure~\ref{fig:cactusC} depicts the cactus plot obtained by posing the query \qline{Extract CactusPlot From * Where hasRegex}. We obtained this plot in roughly $3$ minutes. The cactus plot shows the cumulative time in seconds taken by each solver on all cases in increasing order of runtime. Solvers that are further to the right and closer to the bottom of the plot have better performance. The plot itself shows that \cvc{} seems to implement the most successful algorithm when targeting regular-membership predicates w.r.t. to the analysed instances and embedded solvers.

\medskip

\noindent We have presented here two out of many different possibilities of leveraging information out of a set of benchmarks. Another straightforward use case is the development of a learning algorithm which detects the best solver, from a given set, for each instance by simply extracting features where each solver has its strength. Thus, we could obtain decision trees guiding the selection of solvers on certain data, according to some numerical features of the instance, which can be extracted with our tool.

\section{Conclusion and Future Work:}
In this paper, we have introduced a benchmark analysis framework called \toolname{} to analyse string constraints and string solvers. Our toolbox provides a query language allowing the exploration of a custom benchmark set. \toolname{} provides several useful functions, predicates, and extractors for straight use, within custom queries, and we are continuously working towards enriching the current toolbox with more such operations. Other natural directions of development are to also offer built-in coverage for more general theories (e.g., including the closely related theory of bit-vectors), which are currently treated as generic, as well as to offer good mechanisms for debugging, logging, and diagnostics, especially in the context of user-defined functions. Additionally, our goal is to speed up all sorts of queries, e.g. by smartly combining predicates using the \textsc{SQLite} database or using pre-calculated data more effectively.


\newpage 

\bibliographystyle{splncs04}
\bibliography{bib}


\newpage

\appendix
\section{Appendix}
For clarity, we restate here the link to \toolname{}: \toolWeb{}.

\noindent The bibliographic references in the Appendix refer to the list at its end (and they are given in a different style to avoid confusions).

\subsection{\qlang{} predicates, functions, and extractors}
\label{appendix:qlang}
In our implementation of \toolname{}, a predicate is based on an interface in \texttt{smtquery.smtcon.exprfun} which corresponds to collecting data for the newly defined predicate. After providing a name and a version number, the user implements an \texttt{apply}- and a \texttt{merge}-function as mentioned before. The \texttt{apply}-function receives an AST expression and a pointer to previously calculated data and performs requested modifications to the data. Since the \texttt{apply} function computes the information bottom-up within our AST, the user also provides a neutral element for this computation, which might be an empty dictionary, an empty list, or simply an integer. The interfaces also requires the implementation of a \texttt{merge}-function which aims to combine the data received from children-expressions within the AST in a node. 
Once this information gathering interface is implemented, we register the application within \texttt{smtquery.intel.plugin.probes.intels} by providing a unique identifier which points to a tuple consisting of the function-implementations and the neutral element. Further, we register our predicate at \texttt{smtquery.intel.plugin.probes.predicates} providing a unique name and the predicate. Afterwards, the name is directly usable within our query language. 

The \texttt{apply}-function, primarily allowing modifications of an AST, requires the implementation of a base-class defined within \texttt{smtquery.apply}. We provide a name and the expected behaviour, making sure to return an internal SMT-LIB object. After the successful implementation, we register this new class within our \texttt{PullExtractor} by providing its name. Again, afterwards, the \texttt{apply}-function is immediately usable within the query language.

The extractor allows exporting potentially modified benchmarks in an own format. \toolname{} 
allows either printing the converted data directly to the terminal or redirecting it to a file using \texttt{smtquery.ui.Outputter}. To name a few examples, we might want to translate the benchmarks into a different format, export some plot, or obtain a modified SMT-LIB instance. To implement an extractor, we proceed similarly to the previously seen \texttt{apply}-function and implement a simple class within \texttt{smtquery.extract}. We provide a name and a function preforming the export based on our AST using the \texttt{Outputter}. We again register our new extractor to the \texttt{PullExtractor} by simply providing its class name, allowing us to use it in the query language. Currently, all data exported by our extractors is stored in a seperate folder in \texttt{output} located in the root of \toolname{}.

\subsection{Using \toolname{}}
\label{appendix:usingBASC}
In this section, we explain the basic commands of \toolname{} and showcase some applications of \toolname{} ranging from simple to more sophisticated experiments. This information is also available on our website.

\toolname{} provides a single executable located at \texttt{bin/smtquery} allowing to access all features of our toolbox by positional arguments. We run \toolname{} by executing \texttt{python3 bin/smtquery} in the root folder of the project.  In the following, we list the key arguments while more arguments are explained using the help command.
\begin{enumerate}
   \item \texttt{initdb}: initializes a fresh database containing all instances stored in the file system at \texttt{data/smtfiles}.
  \item \texttt{updateResults}: runs all available SMT-solvers on all registered benchmarks and stores the obtained results.
  \item \texttt{allocateNew}: iterates through the file system and links new benchmark set within the database.
  \item \texttt{qlang}: invokes an interface to pose queries using \qlang{}.
  \item \texttt{smtsolver}: runs an smt-solver on a particular instance, e.g. \texttt{smtsolver CVC5 woorpje track01 01\_track\_1.smt} runs \cvc{} on instance \texttt{01\_track\_1.smt} of track \texttt{track01} of the \texttt{woorpje} benchmark set. 
\end{enumerate}

The next paragraphs list some of the currently implemented predicates, functions, and extractors. The tool is currently under heavily development. Stated today we offer the following predicates, functions and extractors which will be extended continuously. We plan to offer a shared platform to exchange custom implementations of the aforementioned tools.

\paragraph{Predicates.}
To be used within the \qline{Where} part of the query. All predicates can be combined using the common logic connectives, e.g. and, or, and not.
\begin{description}
  \item{\qline{hasWEQ}:} filters to all instances which contain word equations. 
  \item{\qline{hasLinears}:} filters to all instances which contain linear length constraints.
  \item{\qline{hasRegex}:} filters to all instances which contain regular membership predicates.
  \item{\qline{isSimpleRegex}:} filters to all instances which are of the simple regular expression fragment (see \cite{words21}).
  \item{\qline{isSimpleRegexConcatenation}:} filters to all instances which are of the simple regular expression fragment with concatenation (see \cite{words21}).
  \item{\qline{isUpperBounded}:} filters to all instances where the syntax of the formula allows obtaining a length upper bound for each string variable.
  \item{\qline{isQuadratic}:} filters to all instances where each string variable is occurring at most twice.
  \item{\qline{isPatternMatching}:} filters to all instances which only contain word equations of the kind $\mathsf{x} \doteq \alpha$ where $\mathsf{x}$ is a variable not occurring anywhere else in the present formula and $\alpha$ is a string (potentially containing variables other than $\mathsf{x}$).
  \item{\qline{hasAtLeast5Variables}:} filters to all instances containing a least 5 string variables.
    \item{\qline{isSAT($s$)}:}  filters all instances where $s \in \{\cvc,\textsc{Z3Str3},\textsc{Z3Seq}\}$ declared satisfiable.
  \item{\qline{isUNSAT($s$)}:} filters all instances where $s \in \{\cvc,\textsc{Z3Str3},\textsc{Z3Seq}\}$ declared unsatisfiable.
  \item{\qline{hasValidModel(s)}:} filters all instances where $s \in \{\cvc,\textsc{Z3Str3},\textsc{Z3Seq}\}$ returned SAT with a valid model.
    \item{\qline{isCorrect(s)}:} filters all instances where $s \in \{\cvc,\textsc{Z3Str3},\textsc{Z3Seq}\}$ returned SAT with a valid model or UNSAT was returned by the majority of used solvers.
     \item{\qline{isFaster(s1,s2)}:} filters all instances where $s1,s2 \in \{\cvc,\textsc{Z3Str3},\textsc{Z3Seq}\}$ and $s1$ determined some result quicker than $s2$. 
\end{description}
\paragraph{Functions.}
To be used within the \qline{Apply} part of the query.
\begin{description}
  \item{\qline{Restrict2WEQ}:} removes all other predicates than word equations.
  \item{\qline{Restrict2Length}:} removes all other predicates than linear length constraints.
  \item{\qline{Restrict2RegEx}:} removes all other predicates than regular expression membership queries.
  
  \item{\qline{RenameVariables}:} renames all variables to a standard format (i.e. \qline{str01, int01}).
    \item{\qline{DisjoinConstraints}:} splits and-concatenated boolean constraints into separate assertions.
    \item{\qline{ReduceNegations}:} shortens sequences of not, keeping the original polarity.
    \item{\qline{EqualsTrue}:} simplifies constraints comparing boolean expressions to true.
\end{description}
\paragraph{Extractors.}
To be used within the \qline{Extract} part of the query.
\begin{description}
  \item{\qline{MatchingPie}:} exports result as a pie chart.
  \item{\qline{CactusPlot}:} export result as a cactus plot.
  \item{\qline{SMTPlot}:} exports the instances visualized as tree diagram.
  \item{\qline{VarDepPlot}:} exports the dependency plots of all instances.
  \item{\qline{ResultsTable}:} prints the results in terminal.
  \item{\qline{SMTLib}:} exports the resulting instances as SMT-LIB files.
  \item{\qline{Count}:} prints matching instances count and distribution.
    \item{\qline{InstanceTable}:} prints the matching instances and solver’s results.
\end{description}

\subsection{Further Examples}

Next, we show some examples of benchmark analysis, realizable by our~tool.\looseness=-1

\smallskip
\emph{\S\ A variable dependency analysis.} To speed up the solving process for a particular string constraint, one might be interested in splitting a formula into multiple, independent sub-formulae. A relatively na\"ive way of splitting a formula is to determine whether the parts of the input formula in which each variable occurs, and see how they overlap. We can use \toolname{} to visualize the interactions between variables within a formula, using the AST data structure. Whenever an AST is built starting from an SMT-LIB file, we store the variable occurrences for each node, which is actually the only information we need to build a simple variable-dependency graph. So, we have to define the generation of the corresponding plot by implementing the aforementioned class interface in \texttt{smtquery.extract}, call it \texttt{VarDepPlot}, and register it, after implementing the required logic in the \texttt{PullExtractor}. Now we can simply ask a query, e.g. \qline{Extract VarDepPlot From * Where hasWEQ}, creating a variable-dependency plot for each registered instance which contains at least a single word equation. 

\begin{figure}[t]
  \begin{center}
    \begin{tikzpicture}[->,>=stealth',shorten >=1pt,auto,node distance=0.75cm,semithick]
      \node[]   (0)  at (0,0)               {\includegraphics[width=\textwidth]{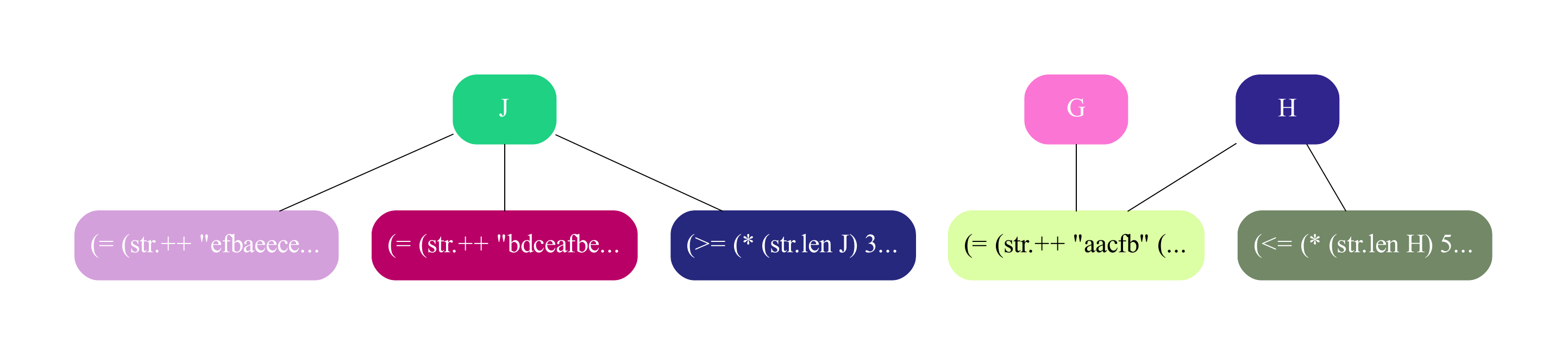}};
    \end{tikzpicture}
  \end{center}
  \vspace*{-1cm}
\caption{Example variable dependency plot.}
\vspace*{-0.5cm}
\label{fig:varplot}
\end{figure}

Posing the query, e.g., have the instance of Listing~\ref{lst:smtexxample2} leading to the plot of Figure~\ref{fig:varplot} where variables $G,H$ and $J$ are top nodes and the corresponding assertions are to bottom nodes. 

\begin{lstlisting}[caption={Instance for variable dependency plot},label=lst:smtexxample2,captionpos=b]
(set-logic QF_S)
(declare-fun H () String)
(declare-fun G () String)
(declare-fun J () String)
(assert (= (str.++  "aacfb" G "abdeddaaa")  (str.++  "aacfbdffebaaaaac" H "aaa") ))
(assert (= (str.++  "efbaeecedaaecfceffaffaedfcebcf" J "aeaadcbe")  (str.++  "e" J "aeecedaaecfceffaffaedfcebcf" J "aeaadcbe") ))
(assert (= (str.++  "bdceafbededddcfcacffdeaefcfa" J "dbabcdebee")  (str.++  "bdceafbededddcfcacffdeaefcfa" J "dbabcdebee") ))
(check-sat)
\end{lstlisting}

The edges in the figure indicate the presence of a variable, allowing us to split the instance accordingly.

\begin{figure}[t]
  \begin{center}
    \begin{tikzpicture}[->,>=stealth',shorten >=1pt,auto,node distance=6cm,semithick]
      \node[]   (0)  at (0,0)               {\includegraphics[width=6cm]{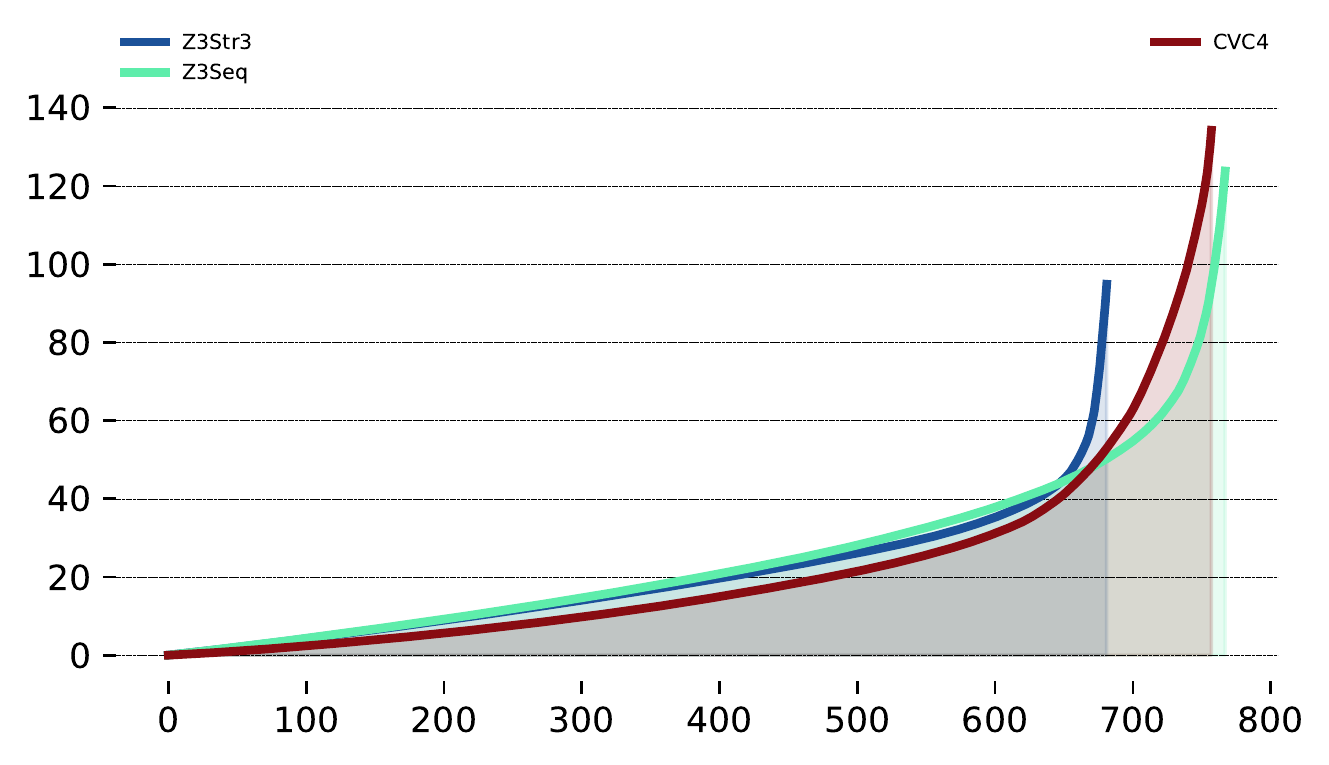}};
      \node[right of=0,text width=5cm]   (0) {\begin{lstlisting}[label=lst:cactus,captionpos=b,basicstyle=\ttfamily\tiny,numbers=none]
                    Z3Str3    Z3Seq     CVC4
----------------  --------  -------  -------
SAT               512       604      594
UNSAT             170       164      164
Unknown            13         0        0
Timeout           114        41       51
Crash               0         0        0
Time w/o Timeout   97.5154  123.816  134.242
Total Time        669.859   329.158  389.631
\end{lstlisting}};
    \end{tikzpicture}
  \end{center}
  \vspace*{-1cm}
\caption{Cactus plot and terminal results for an example query.}
\label{fig:cactusS}
\vspace*{-0.5cm}
\end{figure}

\emph{\S\ Analyzing the performance of a string solver.} To review the performance of a particular solver, one is usually interested in getting a comparison w.r.t. other solvers. \toolname{} allows exporting a summary table and the commonly used cactus plot to compare string solvers on benchmarks. For instance, we want to see whether \cvc{} is performing well enough on the \woorpje{} benchmark set. First, we need to trigger \cvc{} on the respective benchmark by executing, e.g., \qline{Select Name From woorpje where isSAT(CVC5)}. Then, we can obtain a summary table by posing the query \qline{Extract ResultsTable From woorpje} and a cactus plot by simply changing the extractor asking the query \qline{Extract CactusPlot From woorpje}. The results are visualized in Figure~\ref{fig:cactusS}.

\emph{\S\ Modifying Instances.} Analyzing the real-world benchmarks w.r.t. to a particular type of constraints can be achieved by simply neglecting all others. For example, to analyze the structure of the occurring word equations, one may simply pose the query \qline{Extract SMTLib From * Where hasWEQ Apply Restrict2WEQ} to obtain cleaned SMT-LIB files. Naturally, also in this case, we can use any extraction function implemented to acquire the data which we are interested in.

\emph{\S\ Finding and analyzing sub-theories.} In \cite{words21} we have analyzed a large set of benchmarks w.r.t. regular expression membership queries. This kind of queries plays a central role in verifying security policies, by allowing to restrict the set of possible input strings by a regular language \cite{cav21}. The inspection of the respective benchmarks was performed using a sequence of different handcrafted scripts, restricted to a particular use case. \toolname{} provides the means to easily extract this data by simply defining predicates analyzing the regular languages (i.e., regular expressions) occurring in the benchmarks. For example, to gather all instances solely containing regular membership constraints asking whether a string without variables or a single variable is a member of a regular expression without complement or intersection is achieved by posing the query \qline{Select Name From * Where isSimpleRegex}. The key difference is that the definition of the particular predicate is much simpler, due to the extendable structure of \toolname{}. As such, we can now simply combine the acquired information with newly developed predicates. Since this analysis lead to a well performing algorithm, presented in \cite{cav21}, we are optimistic that our tool can be used to extract such relevant data, ultimately leading to better techniques in the area of solving string constraints.

\emph{\S\ Analyzing the structure of instances.} \toolname{} also offers the possibility of a more in-depth analysis of the (syntactic-)structure of the instance. For instance, knowing that all string variables occurring in a formula are subject to constant-length upper bounds allows us to rephrase the problem as a constraint satisfiability problem over finite domains, and ultimately may lead to faster solutions for it. To extract a list of instances having only length-upper-bounded variables, we can pose the query \qline{Select Name From * Where isUpperBounded}. The predicate analyzes the syntax of the constraints and extracts relevant information.
Another interesting aspect, which can potentially lead to a better choice of an algorithm for solving a particular instance, is the analysis of its combinatorial structure. For example, if we know that each variable is occurring at most twice inside a formula, or that each word equation is of the form $\mathsf{x} \doteq \alpha$ where $\mathsf{x}$ is a variable not occurring in $\alpha$ nor anywhere else in the formula, we can use customized solving techniques, and solve the instances more efficiently. Obtaining such information is done by using the predicate \qline{isQuadratic}, resp.  \qline{isPatternMatching}.

\end{document}